\documentclass{aa}
\input epsf.sty
\usepackage{graphics}
\usepackage{times}

\newcommand{\lsi}    {LS~I+61$^{\circ}$303}
\newcommand{\Ha}     {H$\alpha$}
\newcommand{\4}      {$\sim$4~yr}
\newcommand{\ltsima} {$\; \buildrel < \over \sim \;$}
\newcommand{\simlt}  {\lower.5ex\hbox{\ltsima}}            
\newcommand{\gtsima} {$\; \buildrel > \over \sim \;$}
\newcommand{\simgt}  {\lower.5ex\hbox{\gtsima}}            

\begin{document}

\thesaurus{06     
        (08.09.2  
         13.25.5  
         13.18.5  
         08.05.2  
                  ) }

  \title{ First correlation between compact object and circumstellar disk
  in the Be/X-ray binaries   }


  \author {  R.~Zamanov\inst{1,2}  \and  J.~Mart\'{\i}\inst{1} }

  \offprints { J.~Mart\'{\i},\ \ e-mail: jmarti@ujaen.es}

  \institute{ Departamento de F\'{\i}sica, Escuela Polit\'ecnica Superior,
              Universidad de Ja\'en, C/ Virgen de la Cabeza, 2, E--23071 Ja\'en, Spain
  \and National Astronomical Observatory Rozhen,
              P.O.Box 136, BG--4700 Smoljan, Bulgaria }

\date{Received 24 February 2000 / Accepted 2 May 2000 }

\maketitle

\markboth{Zamanov \& Mart\'{\i}: \lsi\ -- Correlation between H$\alpha$\ and radio emission }{}

\begin{abstract}

A remarkable correlation between the \Ha\ emission line 
and the  radio behaviour of \lsi\ over its \4 modulation  
is discovered.  The radio outburst peak is shifted 
by a quarter of the \4\ modulat ion period (about 400 days)
with respect to the equivalent width of the \Ha\ emission line
variability. 
The onset of the \lsi\ radio outbursts varies in phase with the changes of 
the \Ha\ emission line, at least during the increase of 
\Ha\ equivalent width.
This is the first clear correlation 
between the emission associated to the compact object and the Be circumstellar disk
in a Be/X-ray binary system.


\end{abstract}

\section{Introduction}

The Be/X-ray binaries are the major subclass of massive X-ray binary systems 
in which
a neutron star accretes material from the wind of an early type Be star.
The Be stars are known to exhibit emission in the Balmer lines and infrared excess,
which are attributed to the presence of cool circumstellar disk.
Correlation between the changes of the Be circumstellar envelope 
and the emission of the compact object can be expected, as a result of
the compact object interaction with the surrounding matter.
However, no clear correlation has been detected till now -- only loose 
correlations between the optical/infrared properties of the Be 
circumstellar disks and the  X-ray emission of the neutron star have 
been reported to exist (e.g. Corbet et al. 1985; Coe et al. 1994; 
Negueruela et al. 1998). 

\object{\lsi} (\object{V615~Cas}, \object{GT~0236+610}) is a radio emitting 
X-ray binary which exhibits radio outbursts every $26.5\:$d.
The radio outburst peak and the outburst phase  
are known to vary over a time scale of \4 (Gregory et al. 1989; 
Gregory, 1999). 
Hereafter, we will use the latest values reported and we will refer to 
these  radio periods as
$P_{1}=26.4917$~d and $P_{2}=1584$~d. Phase zero for both 
has been set at JD2443366.775 (Gregory, Peracaula \& Taylor, 1999).  
The  $26.5$~d period is believed to be the orbital period.
The \4\ modulation has been discovered on the basis of continued radio monitoring.
Both relativistic jet precession or cyclic
variability in the Be star envelope  have been proposed as 
a possible origin of the long term modulation 
(Paredes, 1987; Gregory et al. 1989), with the second
interpretation being the most likely one.
This suggestion is supported by the fact that the \Ha\ emission line
varies on the same (4 yr) time scale (Zamanov et al. 1999).
However, these authors were not able to derive what is the connection 
between the radio and \Ha\ parameters.

In this letter we report an intriguing correlation between the synchrotron
non-thermal radio emission, associated with the compact star,
and the Be circumstellar disk visible in the $H\alpha$ emission line.  
This is the first clear connection between the Be circumstellar disk variability
and the emission from the neutron star in the Be/X-ray binaries.

\section{$H\alpha$ and radio observations }

The new $H\alpha$ spectroscopic observations used in this paper 
are obtained with the 2-m RCC telescope of
the Bulgarian National Astronomical Observatory `Rozhen' during the last two years. 
They are analyzed together
with the previously published data (Paredes et al. 1994; Zamanov et al. 1999). 
The \Ha\ parameters which vary with the \4\ modulation are 
the equivalent width of \Ha, EW(\Ha), and the distance between the peaks, 
$\Delta V_{\rm peak}$ (Zamanov et al. 1999). 
  
The radio observations during the previous 6 years were retrieved from the Green Bank 
Interferometer, which is a facility of the USA National Science Foundation 
operated by NRAO in support of the NASA High Energy Astrophysics program. 
From this data, we measured the radio outburst peak flux density
and the beginning of every outburst. As onset of the outburst, we adopted the 
time when the radio flux density achieved
a value equal to 1/3 of the maximum flux observed for every orbital period. 
Whenever possible, the onset of the outburst was measured separately 
from the 2.25 GHz and 8.3 GHz observations and the average value is used
in the analysis.

In terms of the Ejector-Propeller model  of \lsi, 
the outburst onset better represents  the time of the 
transition of the neutron star from propeller to ejector and the 
appearance of the expanding radio emitting plasmon (Zamanov, 1995).

  \begin{figure}[htbp]
     \normalsize
     \vskip 3mm plus 1mm minus 1mm
           \epsfysize=13.2cm
           \centerline{\epsffile{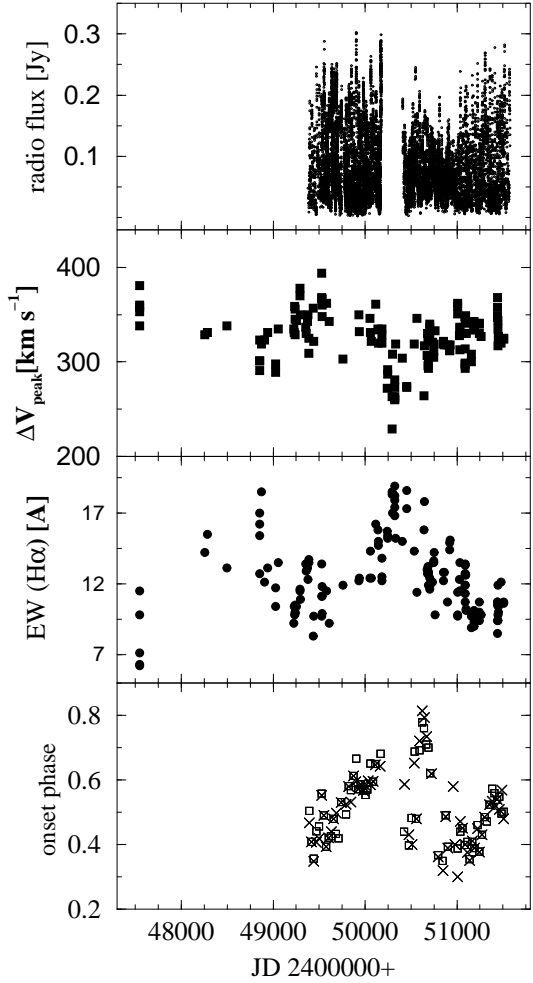}}
  \caption[] { $H\alpha$ and radio observations over the period 1989 - 2000.
  The regular \Ha\ observations cover more than 2500 days. 
  From top to bottom  are plotted the radio flux density at 2.25 GHz, 
  the  $\Delta V_{\rm peak}$ of \Ha, the EW(\Ha), and
  the onset phase of the outburst relatively to the orbital ($P_{1}$) period.   
  In the bottom panel, the squares refer to 2.25 GHz and  ($\times$) - to 8.3 GHz. 
  The traces of the \4\ modulation are visible in all panels. }    
  \label{JDDD}
  \end{figure}
  \begin{figure}[htbp]
     \normalsize
     \vskip 3mm plus 1mm minus 1mm
           \epsfysize=13.2cm
           \centerline{\epsffile{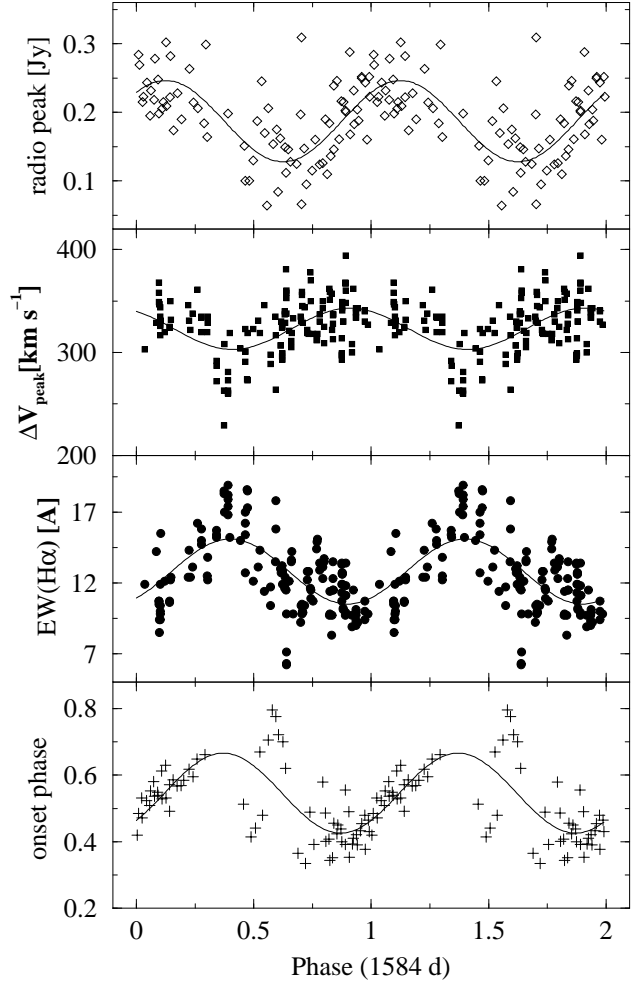}}
  \caption[] { The radio and \Ha\ parameters  folded on  $P_{2}$=1584 day period
  with phase zero at JD2443366.775. The upper panel represents the 
  radio outburst peak flux. The \Ha\ panels are the same as in Fig.\ref{JDDD}. 
  The bottom panel represents the averaged values of the onset phase 
  from 2.25 GHz and 8.3 GHz. 
  }    
  \label{P1584}
  \end{figure}


\section{Relation between circumstellar disk and radio outbursts }

The radio and $H\alpha$ observations obtained during the last years are 
plotted in Fig.\ref{JDDD}. It is clearly visible that all these parameters vary over a time scale 
of \4. Following the predictions of Gregory (1999)  the
radio outburst peak achieved a new maximum at the end of 1999.   
Big changes in the radio
outburst peak flux from values less than 0.150 Jy at  JD2451000  
to values more than  0.250 Jy at JD2451400 are observed (Fig.\ref{JDDD}).  
At the same time, the parameters of the Be circumstellar disk 
$EW(H\alpha)$ and $\Delta V_{\rm peak}$ do not 
exhibit extreme values in their behaviour. This suggests that the
radio peak flux density and the $H\alpha$ parameters do not vary in phase nor in anti-phase. 

To understand this behaviour we plotted all the data folded on the $P_{2}$=1584 d
period. 
Every parameter was fitted separately with a cosine function. Our attempts
to use more complicated functions (e.g. three term truncated 
Fourier fit containing sine and cosine terms with half and quarter 
orbital periods) 
showed that, with this scatter of the points, the 
significance of other terms is negligible and we used simple cosine
waves in the form:  
\hskip 0.3cm  
$y=A+B\:cos(2\pi(\phi +\phi_0))$. 
\hskip 0.3cm 
The best fit parameters are listed in Table~1 and plotted as solid lines
in Fig.\ref{P1584}.

From Table~1 and from Fig.\ref{P1584} the following correlations can be seen:

a) The \Ha\ parameters, 
EW($H\alpha$) and $\Delta V_{\rm peak}$,  
vary in anti-phase. This is a relationship well known for the Be stars, 
reflecting the fact that the the size of the disk grows as the EW(\Ha) increases 
(Hanushik, Kozok \& Kaizer, 1988). 

b) At the same time, a strange phase shift can be seen between the radio outburst peak flux
and the \Ha\ parameters. This shift is about 0.25 in phase. So, if the parameters
really vary as simple cosine functions on average, this means that the radio peak varies
with the first derivative of the $H\alpha$ emission line parameters.

c) The beginning of the radio outburst is 
in phase, or only slightly shifted, relative to the \Ha\ variability,
at least on the ascending branch.

Ray et al. (1997) have shown that the outburst peak is slowly shifting in 
phase during the time interval  from JD2449400 to JD2450100. In the same  
interval  the outburst
peak flux achieved maximum values and began to decline. This is one more 
confirmation that the variability of the outburst peak and the time of the 
outburst are phase shifted one to another over the 1584 day period.  

Although the scatter of the points in Fig.\ref{P1584} is considerable,
all evidences (the cosine fit, the behaviour observed by Ray et al. (1997),  
the behaviour during the last radio outburst maximum) show that the radio
outburst peak  is shifted relatively to the behaviour of the circumstellar disk, 
although the start of the radio outburst is in phase (or almost in phase)
with the circumstellar disk changes.

\section{Discussion}


\begin{table}
\caption[]{ Fitted parameters, $\;\; y=A+B\:cos(2\pi(\phi +\phi_0))$.}

\begin{tabular}{c@{\hspace{1ex}}l@{\hspace{1ex}}cccc}
\hline
 parameter & A & B & $\phi_{0} $  \\ 
\hline
                                  &                  &                 &                 & \\
Radio peak [Jy]                   & 0.187$\pm$0.003  & 0.059$\pm$0.003 & 0.125$\pm$0.010 & \\
Onset phase$^*$                   & 0.548$\pm$0.005  & 0.12$\pm$0.01   & 0.62$\pm$0.01   & \\
EW($H\alpha$) [\AA]             & 12.8$\pm$0.01    & 2.3$\pm$0.1     & 0.60$\pm$0.01   & \\
$\Delta V_{\rm peak}$ [km s$^{-1}$]   & 323$\pm$1    & $-20\pm$2       & 0.59$\pm$0.01 & \\
                                  &                  &                 &                 & \\

\hline  
\end{tabular}

$\;(^*)$ the phase of the beginning of the radio outburst calculated 
relative to the orbital period 

\vskip 0.4cm

\end{table}


A successful modeling of the radio outburst of \lsi\ is
based on the synchrotron radiation from relativistic particles injected
into an expanding plasmon (Paredes et al. 1991). The genesis 
of the plasmon can be a result of the transition of the neutron star from
propeller to ejector state (Zamanov, 1995), or in other words 
from accretion onto the magnetosphere to "young radio pulsar" 
every orbital period.

In such a picture, the start of the outburst will correspond to 
the moment when the neutron star emerges from the denser parts of 
the circumstellar disk. Therefore, the bigger the disk the later 
the outburst can be expected. This can be seen on Fig.\ref{P1584}. 
At phases 0--0.25 we observe increase of the EW(\Ha), decrease of
the $\Delta\:V_{\rm peak}$ and slow shift of the onset of the
outburst. The behaviour of the beginning of the outburst on 
the increasing branch (phases 0--0.25) 
is stable, but on the decreasing branch (phases 0.25--0.5) the scatter
of the points is considerably bigger. The stable behaviour 
at phases 0.25--0.5 is observed twice at about JD~2449800 and 
JD~2451400 (Fig.\ref{JDDD}) so it is unlikely to be a data artifact.
Probably this is a result that, during the disk build-up, 
the increase of the material of the Be disk is feeded
only from one source - the B star equatorial region.
In contrast the  disk-decline can be in two directions - accretion onto the
B star or  slow dissipation outwards. The behaviour of 
the start of the outburst points that the disk build up is a stable 
process and the disk-decline is a more complicated and probably 
unstable process, or may be it suggests formation of structures 
like the double disk observed in \object{X Persei} (Tarasov \& Roche, 1995).    

In context of the propeller-ejector transition the surrounding
matter will basically influence the expansion velocity of the plasmon, 
affecting in this way  the intensity of the radio outbursts. The remaining
plasmon physical parameters (initial magnetic field, injection rate of relativistic
particles, etc.) are not expected to vary significantly from one to another
outburst. 
The expanding plasmon calculations predict that there will be weaker outbursts
for higher expansion velocities. By expanding faster, the energy losses
of the electrons due to the adiabatic expansion are more important and less
electron energy is available to be radiated. In addition, the faster 
decrease of the magnetic field will also contribute to less synchrotron radiation
being produced. Supposing that the plasmon is a result
of the propeller-ejector transition, the expanding plasmon
will appear when the neutron star is receding from the periastron
and the plasmon will expand outside of the \Ha\ emitting disk. 
The size of this disk is about 40-65$\,R_{\odot}$ 
(Zamanov \& Mart\'{\i} 2000) and the apastron separation between components 
is about  150$\,R_{\odot}$ (Hutchings \& Crampton 1981). 
The enigmatic behaviour of the outburst peak flux density (its phase shift
with 0.25 or $\sim$400 days) indicates that the conditions outside the 
\Ha\ disk vary in a different way compared to the changes inside the 
\Ha\  emitting disk. 

The X-ray emission of \lsi\ is observed to exhibit maximum 
every orbital period and the X-ray outburst is shifted relatively to
the radio outburst (Taylor et al. 1996, Harrison et al. 2000). In terms
of the ejector-propeller model, the X-ray maximum is due to the propeller action
and higher mass accretion rate onto the magnetosphere at the periastron passage
(Zamanov \& Zamanova, 1997). 
 In this sense it will be very interesting to see 
what is the behaviour of the X-ray maximum observed in 
the high-energy emission of \lsi\ over the \4 modulation.   

Another possible origin of the \4\ modulation may be the precession of the
Be star. Lipunov \& Nazin (1994) have demonstrated that this value
is in rough agreement with the expected period ($\sim 10^3$d) for  precession of the B
star. The precession of the B star can be expected, because after the
supernova kick the neutron star orbital plane may be different from 
the Be disk plane (e.g. Bradt \& Podsiadlowski, 1995). 
Our attempts to model the behaviour  of the outburst 
phase as a result of the precession are unsuccessful till now but it can be due
to of insufficient data sample, because the systematic radio observations cover
about 6 yr (1.5 periods)  with considerable gaps and scatter inside the data set. 
In this context, it deserves to be
noted  that if the \Ha\ variability is a result of a precessing disk
seen  at different inclination angles, this will imply an inclination
angle  $i>60^\circ$ and a precession angle 
$\Delta i > 6^\circ$ (this estimate is obtained using the values from Table~1 
and assuming everywhere an optically thick in \Ha\ disk).   

To conclude, the behaviour of \lsi\   radio and \Ha\ emission is evidence 
that the picture of the interaction between the neutron star and the circumstellar 
disk in the Be/X-ray binaries is not as simple as generally expected.
We need long series of observations over different wavelengths to better understand
the behaviour of the Be stars and the Be/X-ray binaries.

\begin{acknowledgements}

RZ acknowledges support from Direcci\'on General de Relaciones Culturales
y Cient\'{\i}ficas, Spain. JM acknowledges partial support by 
DGICYT (PB97-0903) and by Junta de Andaluc\'{\i}a (Spain).

\end{acknowledgements}

\end{document}